\documentclass[preprint,tightenlines,eqsecnum,floats,aps,amsmath,amssymb,nofootinbib,prd,showpacs]{revtex4}

\usepackage{amssymb}
\usepackage{stmaryrd}
\usepackage{amsmath}
\usepackage{amsfonts}
\usepackage{mathrsfs}
\usepackage{CJK}
\usepackage{amsmath,amssymb,amsfonts}
\usepackage{graphicx}
\usepackage{subfigure}
\usepackage{color} 
\usepackage{fancyhdr}
\usepackage{hyperref} 
\begin{document}

\title{Solar system constraints of a polymer black hole in loop quantum gravity }

\author{Yun-Long Liu}
\affiliation{Department of Physics, South China University of Technology, Guangzhou 510641, China}
\author{Zhao-Qing Feng}
\affiliation{Department of Physics, South China University of Technology, Guangzhou 510641, China}
\author{ Xiang-Dong Zhang\footnote{Corresponding author. scxdzhang@scut.edu.cn}}
\affiliation{Department of Physics, South China University of Technology, Guangzhou 510641, China}

\date{\today}


\begin{abstract}
A new polymer black hole solution in loop quantum gravity was proposed recently. The difference between the polymer black hole and Schwarzschild black hole is captured by a quantum parameter $A$. In order to get the constraints on parameter $A$, we consider the observational constraints imposed on $A$ by using the Solar System experiments and calculate the deflection of light, Shapiro time delay, perihelion precession and obtain the effects associated with parameter $A$. Moreover, the parameterized post-Newtonian approach of this loop quantum gravity black hole was also carried out. It turns out the tightest constraint on $A$ can be improved to $0<A<4.0\times10^{-6}$.
\end{abstract}
\maketitle
\section{INTRODUCTION}\label{Intro}
Einstein's general relativity(GR) has been proposed for more than a century and still being an exciting and prosperous field.
It has undergone more and more sophisticated tests with the development of tools like atomic clocks, radio telescopes. Currently, the gravity test projects such as the deflection of light, the perihelion advance, and Shapiro time Delay have been carried out in the weak field zone especially in solar system \cite{Confrontation_will_2014}. While in the strong field regime, the gravity tests include the observation of binary pulsars \cite{Binary_taylor_1994, Doublebinarypulsar_yunes_2009,  Testing_seymour_2018}, the direct detections of gravitational waves \cite{Tests_ligo_2016} and the image of the black hole shadow \cite{First_EHT_2019, Gravitational_ehtcollaboration_2020}. Remarkably, the predictions of GR are all consistent with these observations up to now.

However, despite all of these successes, GR is still far from being perfect. From theoretical consideration, unifying GR with quantum mechanics into a consistent quantum gravity theory remains the biggest theoretical challenge to fundamental physics. Among various approaches to quantum gravity, loop quantum gravity (LQG) is notable with its background independence and non-perturbative features \cite{Ro04,Th07,As04,Ma07}.

Inspired by the full LQG, different models of black holes have been constructed to solve the  singularity in the black hole interior. Without loss of generality\cite{Quantum_ashtekar_2005,Quantum_gambini_2014}, in the effective Hamiltonian, one usually replace the $b$ and $c$ which are the components of Ashtekar connection with their holonomies
	\begin{eqnarray}
		b\rightarrow\frac{\sin(\delta_b b)}{\delta_b}, \quad c\rightarrow\frac{\sin(\delta_c c)}{\delta_c},
	\end{eqnarray}
	where parameters $\delta_b$ and $\delta_c$  correspond to the fundamental discreteness of LQG. Under different choices of $\delta_b$ and $\delta_c$,  the current quantization schemes can be divided into three classes \cite{Mass_bodendorfer_2021,Properties_gan_2020}: 1. $\mu_0$-scheme\cite{Loop_modesto_2006,Semiclassical_modesto_2010}, 2. $\bar{\mu}$-scheme \cite{Loop_bohmer_2007,Quantum_alesci_2019}, 3. generalised $\mu_0$-scheme\cite{Quantum_ashtekar_2018,Quantum_ashtekar_2018a,black_olmedo_2017,Loop_corichi_2016}.
	For example, in Ref. \cite{Loop_modesto_2006,Semiclassical_modesto_2010}, the $\mu_0$-scheme is studied where $\delta_b$ and $\delta_c$ are considered as the constants.
Recently, an attractive polymerized black hole \cite{Effective_bodendorfer_2019,Mass_bodendorfer_2021} solution that corresponds to a specific $\bar{\mu}$-scheme is constructed with
	\begin{eqnarray}
		\delta_b=\pm \frac{4 \lambda_j}{\gamma \left|p_b \right|},\quad \delta_c=\pm \frac{8 \lambda_k}{\gamma \sqrt{\left|p_c \right|} }.
	\end{eqnarray} where $p_b$ and $p_c$ are conjugate pairs correspond to $b$ and $c$, $\lambda_j$ and $\lambda_k$ are  the polymerized constants that related to the inverse Planck curvature and Planck length after rescaling of the fiducial cell\cite{Effective_bodendorfer_2019}, respectively.
 This solution results in quantum extensions
of the Schwarzschild black hole. The classical singularity of the Schwarzschild black hole has been resolved by connecting the black hole region with the white hole region through a bounce. Some aspects of this Schwarzschild-like metric have already been studied. For example, the thermodynamic properties of the effective polymer black hole and the corresponding quantum corrections as functions of black and white hole masses have been investigated in Ref. \cite{Quantum_mele_2021}. In Ref. \cite{Gravitational_fu_2021}, the influences of the quantum effects on the weak and strong bending angles of light rays have been studied. More investigations on the polymerized black holes can be found in Ref. \cite{consistent_bouhmadi-lopez_2020,Properties_gan_2020,Liu22}.

Moreover, the rotating black holes are typically found in nature,
In recent Ref. \cite{Testing_brahma_2021}, the authors found a rotational solution of loop quantum gravity black hole (LQGBH) by the Newman-Janis algorithm. Then they investigate the test of loop quantum gravity by the shadow cast of the rotating black hole as well as other experiments. And they obtain a constraint on the quantum correction parameter $A$ as $0<A<7.7\times10^{-5}$. Note that the more precise experiment constraints on the quantum parameter may help us to get a deeper understanding of LQG and consider that the gravitational experiment in the Solar system provides very high precision \cite{Observational_zhu_2020}. In this paper, we will analyze the influence of the polymerized spacetime solution of the classical observations in the Solar system such as the light deflection, Shapiro time delay, perihelion advance, and the Parameterized Post-Newtonian(PPN) approach. In addition, in this paper, we ignore the effects of the angular momentum of the spacetime, the reason for this is two folds. On one hand, although the rotation of the Sun or the Earth can affect the Shapiro time delay and the light deflection, and have a further impact on the constraint upper bound of parameter $A$. However, the impact is very weak and can be safely ignored. On the other hand, the data from the MESSENGER mission \cite{Precession_park_2017} or LAGEOS II \cite{Accurate_lucchesi_2010} has already taken into account and removed the rotational effect(also named ``Lense-Thirring effect''). Therefore, in this paper we need only focus on the non-rotating quantum corrected space-time.

The structure of this paper is organized as follows.
In Sec.\ref{LQGBH}, we give a brief introduction to review the metric of LQGBH.
Then in Sec.\ref{motion}, we obtain the conserved quantities and the motion equations of a test body in the nonrotating loop quantum gravity spacetime.
We study the observational tests including deflection of light, Shapiro time delay and perihelion precession as well as the PPN approach in Sec.\ref{Tests}, and we obtain the constraints on the dimensionless parameter $A$.
The main conclusions and some discussions are given in Sec.\ref{conclusion}.

\section{Nonrotating loop quantum gravity Black hole}\label{LQGBH}
Starting with the non-rotating loop quantum gravity black hole (LQGBH) \cite{Effective_bodendorfer_2019, Mass_bodendorfer_2021, Testing_brahma_2021},
the Schwarzschild-like metric in the loop quantum gravity reads
\begin{eqnarray}
	\label{mLQG}
	ds^2= -8 A M_b^2 \mathcal{A}(r)  dt^2+\frac{dr^2}{8 A M_b^2 \mathcal{A}(r)}+ \mathcal{B}(r) (d\theta^2 + \sin^2(\theta) d\phi^2).
\end{eqnarray}
The metric functions $\mathcal{A}(r)$ and $\mathcal{B}(r)$ are given by
\begin{eqnarray}
	\mathcal{A}(r) &=& \frac{1}{\mathcal{B}(r)} \left(\frac{r^2}{8 A  M_b^2}+1\right) \left(1-\frac{2 M_b}{\sqrt{8 A M_b^2+r^2}}\right),\\
	\mathcal{B}(r) &=&\frac{512 A^3 M_b^4 M_w^2+\left(\sqrt{8 A M_b^2+r^2}+r\right)^6}{8 \sqrt{8 A M_b^2+r^2} \left(\sqrt{8 A M_b^2+r^2}+r\right)^3},
\end{eqnarray}
where $M_b$ and $M_w$ is the mass of asymptotically Schwarzschild black hole and white hole respectively, and the dimensionless parameter  $A$ is defined by $A=(\lambda_k/M_b M_w)^{2/3}/2$. Notice that the parameter $\lambda_j$ will be eliminated during  fixing the integration constants and introducing  $M_b$ and $M_w$  for solving  the  effective equations\cite{Effective_bodendorfer_2019}. Hence, $\lambda_j$ doesn't appear in the metric \eqref{mLQG}.

Without loss of generality, we consider the most interesting and meaningful scheme that $M_b=M_w=M$ \cite{Effective_bodendorfer_2019, Mass_bodendorfer_2021,Testing_brahma_2021}. It displays a symmetric spacetime  reflecting at the transition surface ($r=0$).
In this scheme, the metric functions $\mathcal{A}(r)$ and $\mathcal{B}(r)$ reduce to
\begin{eqnarray}
	\mathcal{A}(r) &=& \frac{1}{\mathcal{B}(r)} \left(\frac{r^2}{8 A  M^2}+1\right) \left(1-\frac{2 M}{\sqrt{8 A M^2+r^2}}\right),\\
	\mathcal{B}(r) &=& 2 A  M^2+r^2.
\end{eqnarray} This metric solves the interior singularity of black hole. While the positive solutions of $\mathcal{A}(r) = 0$ corresponds to the black hole horizon of LQGBH as
\begin{eqnarray}
	r_{+} =  2 \sqrt{M^2-2 A M^2}.
\end{eqnarray}
It should be noted that horizons  will disappear when $A>1/2$. In the $A \to 0$ limit, the expressions of  $8 A M^2 \mathcal{A}(r)$ and $\mathcal{B}(r)$ can be simplified to
\begin{eqnarray}
	8 A M^2 \mathcal{A}(r) &\to&  1 - \frac{2 M}{r},\\
	\mathcal{B}(r) &\to&  r^2.
\end{eqnarray}
Hence, the metric will go back to Schwarzschild case in the classical limit.

\section{motion Equations of test particles} \label{motion}
Let us consider the evolution of a test particle
in the LQGBH spacetime. Note that we have two Killing vectors $K_t^{\mu }=(\frac{\partial}{\partial t})^\mu$ and $K_{\phi }^{\mu }=(\frac{\partial}{\partial \phi})^\mu$. Then the corresponding conserved quantities can be obtained as
\begin{eqnarray}
	\label{E1}
	E&=&-p_{\mu} K_t^\mu = - p_t = -m g_{t t} \dot{t} , \\
	\label{J1}
	J&=& p_{\mu} K_\phi^\mu = p_\phi = m  g_{\phi  \phi } \dot{\phi},
\end{eqnarray}
where the point $\dot{\ }$ represents the covariant derivative with respect to proper time or affine parameter $\tau$ and $m$ is the mass of the test body. $E$ and $J$ denote the conserved energy and the angular momentum of the test body, respectively.
 In addition, the Hamilton-Jacobi equation reads
	\begin{eqnarray}
		\label{HJ}
		\frac{\partial S}{\partial \tau}+H=0,
	\end{eqnarray}
	where $S$ and $H=g_{\mu \nu } p^{\mu} p^{\nu}/2m$ is the Jacobi action and the Hamiltonian, respectively.
	Due to the Eq.\eqref{E1} and Eq.\eqref{J1}, the Jacobi action $S$ can be assumed as
	\begin{eqnarray}
		\label{JaS}
		S = J \phi -E t + \frac{k \tau }{2}+r S_r(r)+ S_\theta(\theta).
	\end{eqnarray} Then, substituting the Ansatz \eqref{JaS} into Eq.\eqref{HJ}, the equation can be separated with  Carter constant $Q$ as follows
	\begin{eqnarray}
		\label{SP}
		&&-8 A M^2 \mathcal{A}(r) \mathcal{B}(r)\left(\frac{\partial S_r(r)}{\partial r}\right)^2+\frac{E^2 \mathcal{B}(r)}{8 A M^2 \mathcal{A}(r)}-k \mathcal{B}(r) \notag\\
		=&&\left(\frac{\partial S_\theta(\theta )}{\partial \theta}\right)^2+J^2 \csc ^2(\theta )=Q,
	\end{eqnarray} with massive particles($k\neq0$) and massless particles($k=0$). Taking into account  the relation between the momentum $p^\mu$ and the Jacobi action that $\partial S/\partial x^\mu =p_\mu$, the Eqs. \eqref{E1}, \eqref{J1},  \eqref{SP}, and  the variables with unity mass($\bar{E}=E/m$, $\bar{J}=J/m$, $\bar{Q}=Q/m^2$), equations of motion for test body read
	\begin{eqnarray}
		\label{dt}
		\frac{dt}{d\tau}&=&\frac{\bar{E}}{8 A M^2 \mathcal{A}(r)}, \\
		\label{dr}
		\left(\frac{dr}{d\tau}\right)^2&=&\bar{E}^2 - \frac{8 A  M^2 \mathcal{A}(r)  \bar{Q} }{\mathcal{B}(r)} - 8 A M^2 \mathcal{A}(r) \bar{k},\\
		\label{dthe}
		\left(\frac{d\theta}{d\tau}\right)^2
		&=&\frac{\bar{Q}}{\mathcal{B}(r)^2}-\frac{\bar{J}^2 \csc ^2(\theta )}{\mathcal{B}(r)^2},\\
		\label{dphi}
		\frac{d\phi}{d\tau}&=&\frac{\bar{J} }{\mathcal{B}(r)},
	\end{eqnarray}
	with $\bar{k}=k/m^2$ ( massive particles($\bar{k}=1$) and  massless one($\bar{k}=0$)). In the following, to simplify the equation of motion, we can always choose a new coordinate system so that the initial position and velocity of the test object are in the equatorial plane($\theta_0=\pi/2$, $\dot{\theta}_0=0 \Rightarrow \bar{Q}=\bar{J}^2$).  With the initial conditions,   the solution to the Eq.\eqref{dthe} is $\theta=\pi/2$ and $\dot{\theta}=0$.  For simplicity, the $\bar{\ }$ is eliminated, and we directly use the $E$, $J$ and $Q$ to denote $\bar{E}$, $\bar{J}$ and $\bar{Q}$, respectively.

\section{Observational tests in the  loop quantum gravity}\label{Tests}
Using the motion of the test body that we have discussed above, the classical tests including the deflection of light, Shapiro time delay, perihelion advance, and geodesic precession of the spinning object will be studied in this section.

\subsection{Deflection of Light}
Starting from the motion equation of massless particle ($k=0$), Eqs.\eqref{dr}-\eqref{dphi} implies
\begin{eqnarray}
	\label{dr_dphi}
	\frac{d r}{d \phi }=\left(\frac{d r}{d \tau }\right)/\left(\frac{d \phi }{d \tau }\right) =\sigma \sqrt{\mathcal{B}(r) \left(\frac{\mathcal{B}(r)}{b^2}-8 A M^2 \mathcal{A}(r)\right)},
\end{eqnarray}
where an ``impact'' parameter $b$ represents the ratio of $J$ and $E$ ($b=J/E$) is introduced. Moreover, the parameter $\sigma = \pm 1$ correspond to the  outgoing and ingoing trajectories, respectively.

The light deflection angle can be expressed as
\begin{eqnarray}
	\label{phir0}
\Delta \phi =2 \int_{r_0}^{\infty } \frac{1}{\sqrt{\mathcal{B}(r) \left(\frac{\mathcal{B}(r)}{b^2}-8 A M^2 \mathcal{A}(r)\right)}} \, dr-\pi,
\end{eqnarray}
where $r_0$ is the turning point of the trajectories which is given by $\frac{d r}{d\phi}|_{r=r_0}=0$.
Combine this condition with Eq.\eqref{dr_dphi}, the relation between  impact parameter $b$ and the closest approach $r_0$   reads
\begin{eqnarray}
	b=\sqrt{\frac{\mathcal{B}\left(r_0\right)}{8 A M^2 \mathcal{A}\left(r_0\right)}}.
\end{eqnarray} To simplify the integration in Eq.\eqref{phir0}, we do the transformation
\begin{eqnarray}
	\label{ru}
	r=\frac{r_0}{u}.
\end{eqnarray}
Then the light deflection angle  formula \eqref{phir0} can be reexpressed as
\begin{eqnarray}
	\label{phiu}
	\Delta \phi =2 \int_0^1 \frac{r_0}{u^2
		\sqrt{ \frac{1}{b^2}\mathcal{B}\left(r_0/u\right)
		\left(\mathcal{B}\left(r_0/u\right)-8 A M^2 \mathcal{A}\left(r_0/u\right)  b^2\right)}} \, du-\pi.
\end{eqnarray} In the weak field limit, the magnitude of $\epsilon=M/r_0$ is a small quantity. By replacing $r_0$ in Eq.\eqref{phiu}  with $\epsilon$ and expanding the integrand, the approximations of the integration in terms of $\epsilon$  is  given by
{\small
\begin{eqnarray}
	\Delta \phi =&&2 \int_0^1
	\left(\frac{1}{\sqrt{1-u^2}}+\frac{\epsilon \left(u^2+u+1\right)}{(u+1) \sqrt{1-u^2}}+\frac{\epsilon^2 \left(3 \left(u^2+u+1\right)^2-4 A (u+1)^2 \left(2 u^2+1\right)\right)}{2 (u+1)^2 \sqrt{1-u^2}}\right)
	\, du  \notag\\
	&&-\pi+O\left(\epsilon^3\right).
\end{eqnarray}
} Then one obtains the
deflection angle of the light as
\begin{eqnarray}
	\Delta \phi \approx \Delta \phi _{GR}-4 \pi   \left(\frac{M}{r_0}\right)^2 A \approx \Delta \phi _{GR} \left(1 - \pi A \frac{M_0}{r_0}  \right),
\end{eqnarray}
where $\Delta \phi _{GR}$ is the light deflection in general relativity.

In the solar system, we assume that the closest distance approaching the Sun $r_0$ of the  electromagnetic wave signal is approximately being the radius of the Sun $R_S$. Hence, the effect of parameter $A$ on light deflection can be described as $\Delta \phi _{\text{GR}}  \pi  A M_S/R_S$ with $M_S$ being the mass of the Sun. Using the measurement data of the radio waves deflection among four quasars sources with the Very Long Baseline Array (VLBA) \cite{PROGRESS_fomalont_2009} in the solar system and considering $A>0$, the constraint of the quantum parameter $A$ can be obtained for
\begin{eqnarray}
0 < A < 74.9836 \quad (68\% \, C.L.).
\end{eqnarray}

\subsection{Shapiro time Delay}
In this section, we still consider the case of a massless test body moving in curved spacetime. Considering Eqs. \eqref{dt} and  \eqref{dr} when the parameter $k$ equals to zero ($k=0$), the  differential equation of massless particles between $t$ and $r$  reads
\begin{eqnarray}
	\label{dt_dr}
	\frac{d t}{d r}=\left(\frac{d t}{d \tau }\right) /\left(\frac{d r}{d \tau }\right) =\pm \sqrt{\frac{\mathcal{B}(r)}{64 A^2 M^4 \mathcal{A}(r)^2 \left(\mathcal{B}(r)-8 A b^2 M^2 \mathcal{A}(r)\right)}}.
\end{eqnarray}
Then the time difference of the electromagnetic wave signal moving from  the closest approach point $P_0$ of the Sun to the point $X$  of satellite or planet as
\begin{eqnarray}
	\Delta  t(r_X)=\int_{r_0}^{r_X} \sqrt{\frac{\mathcal{B}(r)}{64 A^2 M^4 \mathcal{A}(r)^2 \left(\mathcal{B}(r)-8 A b^2 M^2 \mathcal{A}(r)\right)}} \, dr.
\end{eqnarray}
where $r_0$ is the radius of the closest approach point $P_0$ and $r_X$ is the radius of the point $X$.
Doing the same transformation as in Eq.\eqref{ru} and expanding the expression of  time $\Delta  t(r_X)$, with the perturbation quantity $\epsilon=M/r_0$ in the weak field, we found that
\begin{eqnarray}
	\Delta  t(r_X)&=&M \int_{\frac{r_0}{r_X}}^1
	\sqrt{\frac{1}{1-u^2}} \left(\frac{1}{u^2 \epsilon}+\frac{(3 u+2) }{u (u+1)}+\frac{\left(-16 A (u+1)^2+3 u (5 u+8)+12\right) \epsilon}{2 (u+1)^2} \right)  \, du\notag\\
	&&+O\left(\epsilon^2\right).
\end{eqnarray}
The time $\Delta t$ up to the sub-leading order can be expressed as
{\small
\begin{eqnarray}
	\Delta  t\left(r_X\right)
	&\approx&\sqrt{r_X^2-r_0^2}+M \left(\sqrt{\frac{r_X-r_0}{r_X+r_0}}+2 \cosh ^{-1}\left(\frac{r_X}{r_0}\right)\right)
	+\frac{1}{2} M \epsilon  \left(\frac{r_X}{r_X+r_0}\right)^{3/2} \notag \\
	&&\times  \left(\frac{\sqrt{r_X-r_0} \left(4 r_X+5 r_0\right)}{r_X^{3/2}}+2 (16 A-15) \left(\frac{r_X+r_0}{r_X}\right){}^{3/2} \sin ^{-1}\left(\frac{\sqrt{r_X-r_0}}{\sqrt{2 r_X}}\right)\right).
\end{eqnarray}
}

Now, we consider the Shapiro time delay by sending an electromagnetic wave signal out from a source (satellite or  Earth) $X$, and  receiving the signal  reflected by another reflection body (satellite or planet) $Y$.  Whether the electromagnetic wave signal passes perihelion (the turning point of the trajectories $P_0$) or not, the calculation of time delay can be divided into two categories, the inferior conjunction case  and superior one.

In the inferior conjunction case, in which the object $Y$ that reflects the radar signal is located between Earth
(or spacecraft, denoted by $X$) and the Sun. The time delay with the effect of the parameter $A$ reads
\begin{eqnarray}\label{timedelayinferior}
	\delta t_I&\approx& 4  \log \left(\frac{r_X}{r_Y}\right) M
	+\frac{2  r_0 \left(r_X-r_Y\right)}{r_X r_Y} M
	+\frac{2 (8 A-6) \left(r_X-r_Y\right)}{r_X r_Y} M^2	 \notag\\
	&=& \delta t^{GR}_I +\frac{16 A \left(r_X-r_Y\right)}{r_X r_Y} M^2,
\end{eqnarray}
where $\delta t^{GR}_I$ referred to as the Shapiro time delay in Einstein's general relativity.

In the superior conjunction case the object $Y$ reflects the
radar signal and the object $X$ is on opposite sides of the Sun. By taking the similar procedures that in Eq \eqref{timedelayinferior}. The time delay in this case is then given by
\begin{eqnarray}
	\label{Std}
	\delta t_S&\approx&4 M + 4 \log \left(\frac{4 r_X r_Y}{r_0^2}\right) M -\frac{2 \left(r_0 \left(r_X+r_Y\right)\right)}{r_X r_Y} M
	 \notag\\
	&&+\left(\frac{16 \pi  A-15 \pi +8}{r_0}-\frac{4 (4 A-3) \left(r_X+r_Y\right)}{r_X r_Y}\right) M^2 \notag\\
	&=& \delta t^{GR}_S +16 A \left( \frac{\pi}{r_0} - \frac{r_X +r_Y}{r_X r_Y} \right) M^2.
\end{eqnarray}

In astronomical measurements of the Cassini experiment, the researchers always measure the relative change of the radar signal frequency in the superior conjunction case, rather than measuring the detail time delays directly. Combining with  the time delay expression as Eq.\eqref{Std} in the superior conjunction case, the relative change of the frequency of the radar signal reads
\begin{eqnarray}
	\delta \nu &=& \frac{d}{dt} \delta t_S = \frac{d}{dt} \left(  \delta t^{GR}_S +16 A \left( \frac{\pi}{r_0} - \frac{r_X +r_Y}{r_X r_Y} \right) M^2   \right) \notag\\
	 &\approx&  \delta \nu^{GR}_S  -\frac{16 A M^2  \pi  r_0'(t) }{r_0^2}.
\end{eqnarray}

According to the Cassini experiment\cite{test_bertotti_2003}, the frequency shift caused by quantum gravity effect is
\begin{eqnarray}
\delta \nu_A \approx \frac{4096 \pi }{729} \left(\frac{M_S}{R_S}\right)^2 v_E A < 10^{-14},
\end{eqnarray}
where $v_E=r_0'(t)$ is the  Earth's average orbit velocity, $M_S$ and $R_S$ are the mass and the radius of the Sun respectively.
By transferring geometric units($c=G=1$) to SI one and use the experiment data, the constraint on parameter $A$ reads
\begin{eqnarray}
	0 < A < 1.27.
\end{eqnarray}

\subsection{Perihelion Advance}
Considering the motion equation for massive particle ($k=-1$). Eqs. \eqref{dr}-\eqref{dphi} imply a differential equation of massive particles between $r$ and $\phi$  as
\begin{eqnarray}
	\label{dr_dphi_m}
	\frac{d r}{d \phi }=\left(\frac{d r}{d \tau }\right)/\left(\frac{d \phi }{d \tau }\right) =\sigma \sqrt{\frac{ \mathcal{B}(r) \left(E^2  \mathcal{B}(r)-8 A M^2 \mathcal{A}(r) \left(J^2+ \mathcal{B}(r)\right)\right)}{J^2}},
\end{eqnarray} where the parameter $\sigma = \pm 1$ corresponds to the  outgoing and ingoing moving, respectively. By doing the standard transformation $u=r_0/r$, Eq.\eqref{dr_dphi_m} can be restructured to
\begin{eqnarray}
	\label{du_dphi}
	\left(\frac{d u}{d \phi }\right)^2=\frac{u^4 \mathcal{B}\left(r_0/u\right) }{r_0^2}
	\left(\frac{E^2 \mathcal{B}\left(r_0/u\right)}{J^2}-8 A M^2 \mathcal{A}\left(r_0/u\right) \left(\frac{\mathcal{B}\left(r_0/u\right)}{J^2}+1\right)\right).
\end{eqnarray} By introducing the small parameter $\epsilon=M /r_0 $, the above equation can be expressed as:
\begin{eqnarray}
	 u''(\phi )+u(\phi) =f,
\end{eqnarray}
where $f$ is the function of $u(\phi)$ as
\begin{eqnarray}
	f &=&\frac{1}{ J^2 \mathcal{I} \epsilon }
	\Big(
	M^2
	+2 \left(2 E^2-5\right) A M^2 \mathcal{I} \epsilon  u(\phi )
	+\left(22 A M^2+3 J^2\right)  \epsilon ^2 u(\phi )^2
	 \notag\\
	&&
	+  8 A \mathcal{I}  \left(\left(E^2-4\right) A M^2-2 J^2\right) \epsilon ^3 u(\phi )^3
	+32 A  \left(2 A M^2+J^2\right)  \epsilon ^4 u(\phi )^4    \Big),
\end{eqnarray}
and
 \begin{eqnarray}
 	\mathcal{I}=\sqrt{8 A \epsilon ^2 u(\phi )^2+1}.
 \end{eqnarray} Next, we use the perturbation method to obtain the approximate solution of the above equation.
Hence, we need to consider the approximations of the differential equation in terms of $\epsilon$ in the weak field where the magnitude of $\epsilon= M / r_0$ is small, and the above differential equation can be truncated as
 \begin{eqnarray}
 	\label{d2uphi}
	u''(\phi )+u(\phi ) - \frac{Mr_0}{J^2 }&=&
	\frac{2 \left(2 E^2-5\right) A M^2 u(\phi )}{J^2}
	+\frac{3  \left(6 A M^2+J^2\right) u(\phi )^2 \epsilon }{4 J^2}
	\notag\\
	&&
	+\frac{ A   \left(\left(E^2-4\right) A M^2-2 J^2\right) u(\phi )^3 \epsilon ^2}{2 J^2}
	+ O\left(\epsilon ^3\right).
\end{eqnarray} The unperturbed solution of Eq.\eqref{d2uphi} can be obtained by solving the equation
\begin{eqnarray}
	\label{und2u}
	u_0''(\phi )+u_0(\phi )-\frac{Mr_0}{J^2 }=0,
\end{eqnarray}
the solution of Eq.\eqref{und2u} reads
\begin{eqnarray}
	\label{u0}
	u_0(\phi )=\frac{Mr_0 }{J^2 } (1 + e \cos (\phi )).
\end{eqnarray} Second, consider the first-order correction $u_1$ of the unperturbed orbit $u_0$. The approximate solution can be expanded as $u\approx u_0+u_1$, where the $u_1$ should satisfy
 \begin{eqnarray}
 	\label{Du1_1}
	u_1''(\phi )+u_1(\phi ) - \frac{Mr_0}{J^2 }&\approx&
	\frac{2 \left(2 E^2-5\right) A M^2 u_0(\phi )}{J^2}
	+\frac{3   \left(6 A M^2+J^2\right)  u_0(\phi )^2 \epsilon}{4 J^2}\notag\\
	&&+\frac{ A   \left(\left(E^2-4\right) A M^2-2 J^2\right) u_0(\phi )^3 \epsilon ^2}{2 J^2}.
\end{eqnarray}
Plugging the Eq.\eqref{u0} into Eq.\eqref{Du1_1}, we can find that
\begin{eqnarray}
	\label{Du1_2}
	u_1''(\phi )+u_1(\phi ) &=&  \sum_{n=0}^{3} A_n \cos^n(\phi),
\end{eqnarray}
where
\begin{eqnarray}
	\label{A_03}
	\mathcal{A}_0&=&\frac{1}{J^8 \epsilon }
	\left(8 \left(E^2-4\right) A^2 M^8+2 A J^2 M^6+\left(2 \left(2 E^2-5\right) A+3\right) J^4 M^4\right),\\
	\mathcal{A}_1&=&\frac{1}{J^8 \epsilon }
	\left(2 e \left(12 \left(E^2-4\right) A^2 M^8-6 A J^2 M^6+\left(\left(2 E^2-5\right) A+3\right) J^4 M^4\right)\right),\\
	\mathcal{A}_2&=&\frac{1}{J^8 \epsilon }
	\left(  3 e^2 M^4 \left(8 \left(E^2-4\right) A^2 M^4-10 A J^2 M^2+J^4\right)   \right),\\
	\mathcal{A}_3&=&\frac{1}{J^8 \epsilon }
	\left(   8 A e^3 M^6 \left(\left(E^2-4\right) A M^2-2 J^2\right)  \right).
\end{eqnarray} Considering the initial conditions
\begin{eqnarray}
	\label{In}
	u_1(0)=0,  \quad  u_1'(0)=0.
\end{eqnarray}
The solution $u_1$ reads
\begin{eqnarray}
	u_1(\phi)=\sum_{n=0}^{3} C_n \cos(n \phi) + S_1 \phi \sin(\phi),
\end{eqnarray}
where
\begin{eqnarray}
	C_0 &=& \mathcal{A}_0+\frac{\mathcal{A}_2}{2},\\
	C_1 &=& -\left(\mathcal{A}_0+\frac{\mathcal{A}_2}{3}-\frac{\mathcal{A}_3}{32}\right),\\
	C_2 &=& -\frac{1}{6} \mathcal{A}_2, \\
	C_3 &=& -\frac{1}{32} \mathcal{A}_3,\\
	S_1 &=& \frac{\mathcal{A}_1 \phi }{2}+\frac{3 \mathcal{A}_3 \phi }{8}.
\end{eqnarray}
It is easy to see that the perihelion advance only depends on the $S_1 \phi \sin(\phi)$ term in $u_1$. Hence, the $\sum_{n=0}^{3} C_n \cos(n \phi)$ term can be ignored, and the approximation solution of $u$ reads
\begin{eqnarray}\label{uphi}
	u(\phi) &\approx&
	\frac{Mr_0}{J^2  }
	+\frac{e Mr_0 }{  J^2}
	\left(\mathcal{P} \sin (\phi )+\cos (\phi )\right)\notag\\
	&=&\frac{Mr_0}{J^2 }
	\left(
	1 + e  \sqrt{1+\mathcal{P}^2}\cos(\phi - \phi_0)
	\right),
\end{eqnarray}
where
\begin{eqnarray}
	\mathcal{P}&=&
	\left(2 E^2 A-5 A+3\right) \frac{ M^2 \phi }{J^2}
	-6 A \left(e^2+1\right) \frac{ M^4 \phi }{J^4}
	+3 \left(E^2-4\right) A^2 \left(e^2+4\right) \frac{ M^6 \phi }{J^6},\\
	\phi_0&=&\frac{\delta \phi _0  }{2 \pi }\phi=\arctan(P).
	\label{phi0}
\end{eqnarray} Now, considering the orbit of the solution \eqref{uphi}, the perihelion radius $r_-$ and aphelion radius $r_+$ of the orbit read respectively
\begin{eqnarray}
	\label{rp}
	\frac{r_0}{r_+}=\frac{(1-e) r_0}{ M} \frac{ M^2 }{J^2},\\
	\label{rn}
	\frac{r_0}{r_-}=\frac{(1+e) r_0}{ M}\frac{ M^2 }{J^2}.
\end{eqnarray} Combining the Eqs.\eqref{rp} and \eqref{rn}, we found that
\begin{eqnarray}
	\frac{M^2}{J^2}=\frac{2 M}{\left(1-e^2\right) \left(r_-+r_+\right)} = \frac{ M}{\left(1-e^2\right) \kappa}  \sim \frac{M}{r_0} = \epsilon,
\end{eqnarray}
where $\kappa:=(r_+ + r_-)/2$ is the semi-major axis of the orbit.
Therefore $M^2/J^2$ and $\epsilon=M/r_0$ have the same order of magnitude. Hence, the angular $\phi_0$ in Eq.\eqref{phi0} can be expanded in terms of $M^2/J^2$, and  the angular shift of  the perihelia per orbit $\delta \phi _0$ reads
\begin{eqnarray}
	\label{Adphi}
	\delta \phi _0 \approx
	2 \pi  \left(3 + \left(2 E^2-5\right) A\right) \frac{ M^2}{J^2} + O\left(\left(\frac{ M^2}{J^2}\right)^2\right).
\end{eqnarray} By solving $ dr / d\phi|_{r_0} = 0$ with Eq.\eqref{dr_dphi_m}, the relation between the energy $E$ and the closest approach $r_0$ reads
\begin{eqnarray}
	E^2 = 8 A M^2 \mathcal{A}(r_0) \left(\frac{J^2}{\mathcal{B}(r_0)}+1\right) = 1 - 2\epsilon + O(\epsilon^2).
\end{eqnarray}
Therefore, in the weak field approximation, the Eq.\eqref{Adphi} can be simplified as
\begin{eqnarray}
	\Delta \phi =\delta \phi _0  \approx \frac{6 \pi  M}{\left(1-e^2\right) \kappa}(1-A + O(\epsilon^1)) \approx  \Delta \phi^{GR} (1-A).
\end{eqnarray} Now, we could get an upper bound of the parameter $A$ by using the observational data.
For the experimental data of the anomalous Mercury perihelion advance from the MESSENGER mission \cite{Precession_park_2017}, the  precession rate of perihelion  caused by the gravitoelectric effect reads
\begin{eqnarray}
	\Delta\phi = (42.9799 \pm 0.0009)''/\text{century}.
\end{eqnarray}
For the motion of Mercury around the Sun, the observed error of anomalous perihelion advance  is $0.0009''/\text{century}$.  The contribution of LQG is expected to be less than the observational error. Therefore, the constraint range of parameters $A$ turns out to be
\begin{eqnarray}
	0<A<2.09 \times 10^{-5}.
\end{eqnarray}

Analogically, using  the perihelion advance experimental data of the LAGEOS satellites that move around the Earth \cite{Accurate_lucchesi_2010}
\begin{eqnarray}
	\Delta \phi = \Delta \phi ^{GR} (1 + (0.28 \pm 2.14)\times 10^{-3}),
\end{eqnarray}
we can obtain the constraint range of parameters $A$ as
\begin{eqnarray}
	0<A<1.86 \times  10^{-3}.
\end{eqnarray}


\subsection{ Parameterized Post-Newtonian(PPN) approach}
Now we are going to calculate the PPN parameters of the Schwarzschild like metric \eqref{mLQG} in the loop quantum gravity and obtain the relation between the PPN parameter and the parameter $A$ that contains the LQG effect.
First, we perform the following transformations \cite{New_rezzolla_2014}
\begin{eqnarray}
	\label{barr}
	r&=&\sqrt{\bar{r}^2-2A M^{2}}.
\end{eqnarray}
The metric \eqref{mLQG} then can be reformulated as
\begin{eqnarray}
	\label{dSt}
	ds^2= -\mathcal{N}^2(\bar{r})  dt^2+\frac{\mathcal{B}^2(\bar{r})}{\mathcal{N}^2(\bar{r})}d\bar{r}^2+ \bar{r}^2 (d\theta^2 + \sin^2\theta d\phi^2),
\end{eqnarray}
where
\begin{eqnarray}
	\mathcal{N}^2(\bar{r})&=&1-\frac{2M^{2}}{\bar{r}^2}(-3A+\sqrt{A}\sqrt{6+\frac{\bar{r}^2}{A M^{2}}}),\\
	\mathcal{B}^2(\bar{r})&=&\frac{\bar{r}^2}{-2A M^{2}+\bar{r}^2}.
\end{eqnarray}
From the transformation \eqref{barr}, we know that $\bar{r}>\sqrt{2 A} M$ or $\bar{r}<- \sqrt{2 A} M$  must be satisfied to insure that $\mathcal{B}^2(\bar{r})>0$.
On the one hand, the metric component of \eqref{dSt} can be expanded in terms of $M/\bar{r}$ as \cite{Metric_johannsen_2011,New_rezzolla_2014}
\begin{eqnarray}
	\label{NPN}
	\mathcal{N}^2(\bar{r})&=&1-2 \frac{M}{\bar{r}} + 6 A \frac{M^2}{\bar{r}^2} + O\left(\left(M/\bar{r}\right)^3\right),\\
	\frac{\mathcal{B}^2(\bar{r})}{\mathcal{N}^2(\bar{r})}&=&1 + 2 \frac{M}{\bar{r}} + O\left(\left(M/\bar{r}\right)^2\right).
	\label{BPN}
\end{eqnarray}
On the other hand,  the PPN approximation of the asymptotic spacetime can be described as
\begin{eqnarray}
	ds^2= -\mathcal{G}^2(\bar{r})  dt^2+\mathcal{F}^2(\bar{r})d\bar{r}^2+ \bar{r}^2 (d\theta^2 + \sin^2\theta d\phi^2),
\end{eqnarray}
where
\begin{eqnarray}
	\label{GPN}
	\mathcal{G}^2(\bar{r})&=&1-2 \frac{M}{\bar{r}} + 2 \left(\beta - \gamma \right) \frac{M^2}{\bar{r}^2},\\
	\mathcal{F}^2(\bar{r})&=&1 + 2 \gamma \frac{M}{\bar{r}}.
	\label{FPN}
\end{eqnarray}
Here $\gamma$ and $\beta$ are the PPN parameters. Comparing equations \eqref{NPN}-\eqref{BPN} and \eqref{GPN}-\eqref{FPN}, we immediately obtain the corresponding equations as follows
\begin{eqnarray}
	\beta - \gamma = 3 A,\quad \gamma=1
\end{eqnarray}
or equivalently
\begin{eqnarray}
	\gamma=1, \quad  \beta  = 3 A+1.
\end{eqnarray}
Next, we consider observational constraints that imposed to PPN parameter $\beta$
by the  MESSENGER mission \cite{Precession_park_2017} which reads
\begin{eqnarray}
	-6.6 \times 10^{-5}<\beta -1<1.2 \times 10^{-5},
\end{eqnarray}
and this immediately in turn implies
\begin{eqnarray}
	0<A<4 \times 10^{-6}.
\end{eqnarray}

\begin{table}[b]
	\caption{\label{SumTab}
		Summary of estimates for upper bounds of the parameter $A$.}
	\begin{ruledtabular}
		\begin{tabular}{lcl}
			Experiments/Observations  &  $A$  &  Datasets												\\
			\hline
			Light deflection	& 74.9836 					& VLBI observation of quasars\cite{PROGRESS_fomalont_2009}	\\
			Time delay			& 1.27						& Cassini experiment\cite{test_bertotti_2003}	\\
			Perihelion advance	& $2.09\times10^{-5}$		& MESSENGER mission\cite{Precession_park_2017}\\
								& $1.86 \times  10^{-3}$	& LAGEOS satellites\cite{Accurate_lucchesi_2010}\\
			PPN approach($\beta =3 A+1$)& $4.0\times10^{-6}$    & MESSENGER mission\cite{Precession_park_2017} \\	
			Shadow of black hole\cite{Testing_brahma_2021}& 0.24					& Shadow of  M87*\cite{First_EHT_2019}		\\
			Tests strong equivalence principle\cite{Testing_brahma_2021}& $7.7\times10^{-5}$& Lunar laser ranging data \cite{Progress_williams_2004}	\\					
		\end{tabular}
	\end{ruledtabular}
\end{table}

\section{CONCLUSIONS}\label{conclusion}
Cosmological and black holes models inspired by
LQG provide elegant solutions to black holes and big bang singularities have been constructed. Along this line, recently, a new polymer black hole model has been proposed \cite{Effective_bodendorfer_2019,Mass_bodendorfer_2021,Testing_brahma_2021}.
In this paper, we study the classical tests of polymerized black hole in effective loop quantum gravity including the light deflection, Shapiro time delay, perihelion advance, and PPN methods. Based on these classical observations, we calculate the influences of the parameter $A$ and then obtain the constraints on the parameter $A$ using the  latest astronomical observations  in the Solar System.

The upper bounds of the parameter $A$ are summarized in Table \ref{SumTab}.
We interestingly observed that the MESSENGER mission gives a nice constraint on the parameter $A$ through perihelion advance as $0<A<2.09\times10^{-5}$. Moreover, the best constraint comes from the PPN method which gives rise $0<A<4.0\times10^{-6}$.

Note that the observations such as the light deflection and Shapiro time delay do not impose tight constraints on parameter $A$.
The reason is that parameter $A$ will only appears at the  ``nonlinearity''($M^2/r^2$) term that related to the PPN parameter $\beta$ in the Eq. \eqref{NPN}. In contrast, the quantum parameter $P$ in Ref. \cite{Observational_zhu_2020} appears at the ``linearity'' term ($M/r$) and  the ``nonlinearity''($M^2/r^2$) term that will relate to the $\gamma$ and effective gravitational ``constant''  $\bar{G}$ in Appendix \ref{Append}.
Moreover, in the MESSENGER mission experiment \cite{Precession_park_2017}, constraints on the PPN parameter $\beta$ and un-normalized solar quadrupole moment $J_2$ are given directly.
However, the constraint of perihelion advance is determined by the estimated uncertainty of other parameters such as $\beta$, $J_2$ , and $\gamma$  from Cassini. Hence, using the constraint on PPN parameter $\beta$ in MESSENGER mission get the better constraints on $A$ than perihelion advance one.

It is worth noting that in 2018, the joint European-Japanese BepiColombo project launched two spacecrafts that will explore Mercury \cite{BepiColombo_benkhoff_2010,New_will_2018}.
Through the BepiColombo spacecrafts, the accuracy of  Mercury's perihelion advance measurements will be further improved by an order of magnitude compared to the MESSENGER mission and the thus tighter constraint on the parameter $A$ will be obtained accordingly. We would like to leave this for future study.

\begin{acknowledgments}
This work is supported by NSFC with No.11775082.

\end{acknowledgments}

\appendix
\section{PPN approach for the self-dual spacetime in loop quantum gravity}\label{Append}

Starting with the self-dual spacetime in LQG \cite{Semiclassical_modesto_2010,Observational_zhu_2020}, the Schwarzschild-like metric reads
	\begin{eqnarray}
		\label{selfM}
		ds^2=-f(r) dt^2  + \frac{dr^2}{g(r)}+ h(r) d\Omega^2,
	\end{eqnarray} where
	\begin{eqnarray}
		h(r)&=&\frac{a_0^2}{r^2}+r^2,\\
		f(r)&=&\frac{(r-r_-) (r-r_+) (r+r_*)^2}{a_0^2+r^4},\\
		g(r)&=&\frac{r^4 (r-r_-) (r-r_+)}{\left(a_0^2+r^4\right) (r+r_*)^2},
	\end{eqnarray} and
	\begin{eqnarray}
		r_-=\frac{2 M P^2}{(P+1)^2},
		\quad 	r_+=\frac{2 M}{(P+1)^2},	
		\quad r_*=\frac{2 M P}{(P+1)^2}.
	\end{eqnarray}
	Assume that $a_0=0$ \cite{Semiclassical_modesto_2010,Observational_zhu_2020}, and expand the function $f(r)$ and $1/g(r)$
	\begin{eqnarray}
		\label{SDstf}
		f(r)&=&1-\frac{2 (P-1)^2}{(P+1)^2} \frac{M}{r}-\frac{8 P \left(P^2-P+1\right)}{(P+1)^4} \frac{M^2}{r^2} + O\left(\left(M/r\right)^3\right),\\
		\frac{1}{g(r)}&=&1+2 \frac{M}{r}+\frac{4 \left(P^2+1\right) }{(P+1)^2}\frac{M^2}{r^2}+O\left(\left(M/r\right)^3\right).
		\label{SDstg}
	\end{eqnarray}
	Comparing equations \eqref{GPN} and \eqref{FPN}  with \eqref{SDstf} and \eqref{SDstg}, the parameter $P$ does not appear in the linear term of Eq.\eqref{SDstg} that corresponding to the PPN parameter $\gamma$ in the Eq.\eqref{FPN}. In order to obtain the relation between PPN parameter and $P$, we must do some transformations. Inspired by Eq.(202) in Ref. \cite{Gravity_will_2016}, we take the transformation as
	\begin{eqnarray}
	\bar{G}= \frac{(P-1)^2}{(P+1)^2} G.
	\end{eqnarray} The function $f(r)$ and $1/g(r)$ can be reformulated as
    \begin{eqnarray}
    	\label{reSDstf}
    	f(r)&=&1-2 \frac{\bar{G} M}{c^2 r}
    	-\frac{8 P \left(P^2-P+1\right)}{(P-1)^4}\frac{\bar{G}^2 M^2}{c^4 r^2}
    	+ O\left(\left(\bar{G} M/(c^2r)\right)^3\right),\\
    	\frac{1}{g(r)}&=&1
    	+2 \frac{(P+1)^2}{(P-1)^2}\frac{\bar{G} M}{c^2 r}
    	+\frac{4 (P+1)^2 \left(P^2+1\right)}{(P-1)^4}\frac{\bar{G}^2 M^2}{c^4 r^2}
    	+ O\left(\left(\bar{G} M/(c^2r)\right)^3\right).
    	\label{reSDstg}
    \end{eqnarray}
 	Comparing equations \eqref{GPN}-\eqref{FPN} with \eqref{reSDstf}-\eqref{reSDstg}, we immediately read off the relation between PPN coefficients and parameter $P$ as
 	\begin{eqnarray}
 	\gamma &=& \frac{(P+1)^2}{(P-1)^2}=1+4 P+8 P^2+O\left(P^3\right),\\
 	\beta &=& \frac{((P-4) P+1) \left(P^2+1\right)}{(P-1)^4}=1-4 P^2+O\left(P^3\right).
 	\end{eqnarray}
 	Using this relation, we can calculation the light deflection, Shapiro time delay and perihelion
 	advance in  the self-dual spacetime directly. For example, perihelion
 	advance per orbit can be expressed as
    \begin{eqnarray}
    	\Delta\phi
    	= \frac{6 \pi  \bar{G} M}{\left(1-e^2\right) c^2\kappa}\left(\frac{1}{3} (2+2 \gamma -\beta )\right)
    	= \frac{6 \pi G M}{\left(1-e^2\right) c^2\kappa} \left(1-\frac{4 P}{3}+O\left(P^2\right)\right).
    \end{eqnarray}
	This replicates the result of Ref. \cite{Observational_zhu_2020}. We find that $\gamma$ and the effective gravitational ``constant''  $\bar{G}$  contribute $P$  to the leading order of light deflection, Shapiro time delay and perihelion advance, while $\beta$ only contributes $P^2$ that can be ignored. On the contrary, in our work, $A$ is  contributed by $\gamma$ in the leading order of  the perihelion advance, and  $A$ only appears in the second order of the light deflection and Shapiro time delay and hence it is easy to understand that they are not of the same order of magnitude.

\bibliographystyle{unsrt}

\begin{thebibliography}{10}
	\bibitem{Confrontation_will_2014}
	C. M. Will,
	The Confrontation between General Relativity and Experiment,
	Living Rev. Relativ. 17, 4 (2014).
	
	\bibitem{Binary_taylor_1994}
	J. H. Taylor,
	Binary Pulsars and Relativistic Gravity,
	Rev. Mod. Phys. 66, 711 (1994).
	
	\bibitem{Doublebinarypulsar_yunes_2009}
	N. Yunes and D. N. Spergel,
	Double-Binary-Pulsar Test of Chern-Simons Modified Gravity,
	Phys. Rev. D 80, 042004 (2009).
	
	\bibitem{Testing_seymour_2018}
	B. Seymour and K. Yagi,
	Testing General Relativity with Black Hole-Pulsar Binaries,
	Phys. Rev. D 98, 124007 (2018).
	
	\bibitem{Tests_ligo_2016}
	B. P. Abbott, R. Abbott, T. D. Abbott, et al. (LIGO Scientific and Virgo Collaborations),
	Tests of General Relativity with GW150914,
	Phys. Rev. Lett. 116, 221101 (2016).
	
	\bibitem{First_EHT_2019}
	K.Akiyama, A. Alberdi, et al. (EHT Collaboration),
	First M87 Event Horizon Telescope Results. I. The Shadow of the Supermassive Black Hole,
	Astrophys. J. Lett. 875, L1 (2019).
	
	\bibitem{Gravitational_ehtcollaboration_2020}
	D. Psaltis, L. Medeiros, et al. (EHT Collaboration),
	Gravitational Test beyond the First Post-Newtonian Order with the Shadow of the M87 Black Hole,
	Phys. Rev. Lett. 125, 141104 (2020).
	
	\bibitem{Ro04}
	C. Rovelli,
	Quantum Gravity,
	Cambridge University Press, 2004.
	
	\bibitem{Th07}
	T. Thiemann,
	Modern Canonical Quantum General Relativity,
	Cambridge University Press, 2007.
	
	
	\bibitem{As04}
	A. Ashtekar and J. Lewandowski,
	Background independent quantum gravity: A status report,
	Class. Quant. Grav. 21, R53 (2004).
	
	\bibitem{Ma07}
	M. Han, W. Huang, and Y. Ma,
	Fundamental structure of loop quantum gravity,
	Int. J. Mod. Phys. D  16, 1397 ,(2007).
	
	\bibitem{Quantum_ashtekar_2005}
	A. Ashtekar and M. Bojowald,
	Quantum Geometry and the Schwarzschild Singularity,
	Class. Quantum Grav. 23, 391 (2005).
	
	\bibitem{Quantum_gambini_2014}
	R. Gambini, J. Olmedo, and J. Pullin,
	Quantum Black Holes in Loop Quantum Gravity,
	Class. Quantum Grav. 31, 095009 (2014).
	
	
	\bibitem{Mass_bodendorfer_2021}
	N. Bodendorfer, F. M. Mele, and J. Munch,
	Mass and Horizon Dirac Observables in Effective Models of Quantum Black-to-White Hole Transition,
	Class. Quantum Grav. 38, 095002 (2021).
	
	\bibitem{Properties_gan_2020}
	W. Gan, N. O. Santos, F. Shu, and A. Wang,
	Properties of the Spherically Symmetric Polymer Black Holes,
	Phys. Rev. D 102, 124030 (2020).
	
	
	\bibitem{Loop_modesto_2006}
	L. Modesto,
	Loop Quantum Black Hole,
	Class. Quantum Grav. 23, 5587 (2006).
	
	\bibitem{Semiclassical_modesto_2010}
	L. Modesto,
	Semiclassical Loop Quantum Black Hole,
	Int J Theor Phys 49, 1649 (2010).
	
	\bibitem{Loop_bohmer_2007}
	C. G. Bohmer and K. Vandersloot,
	Loop Quantum Dynamics of the Schwarzschild Interior,
	Phys. Rev. D 76, 104030 (2007).
	
	\bibitem{Quantum_alesci_2019}
	E. Alesci, S. Bahrami, and D. Pranzetti,
	Quantum Gravity Predictions for Black Hole Interior Geometry,
	Physics Letters B 797, 134908 (2019).
	
	\bibitem{Quantum_ashtekar_2018}
	A. Ashtekar, J. Olmedo, and P. Singh,
	Quantum Transfiguration of Kruskal Black Holes,
	Phys. Rev. Lett. 121, 241301 (2018).
	
	\bibitem{Quantum_ashtekar_2018a}
	A. Ashtekar, J. Olmedo, and P. Singh,
	Quantum Extension of the Kruskal Spacetime,
	Phys. Rev. D 98, 126003 (2018).
	
	\bibitem{black_olmedo_2017}
	J. Olmedo, S. Saini, and P. Singh,
	From Black Holes to White Holes: A Quantum Gravitational, Symmetric Bounce,
	Class. Quantum Grav. 34, 225011 (2017).
	
	\bibitem{Loop_corichi_2016}
	A. Corichi and P. Singh,
	Loop Quantization of the Schwarzschild Interior Revisited,
	Class. Quantum Grav. 33, 055006 (2016).
	
	
	\bibitem{Effective_bodendorfer_2019}
	N. Bodendorfer, F. M. Mele, and J. Munch,
	Effective Quantum Extended Spacetime of Polymer Schwarzschild Black Hole,
	Class. Quantum Grav. 36, 195015 (2019).


	\bibitem{Quantum_mele_2021}
	F. M. Mele, J. Munch, and S. Pateloudis,
	Quantum Corrected Polymer Black Hole Thermodynamics: Mass Relations and Logarithmic Entropy Correction,
	arXiv:2102.04788 [gr-qc] (2021), 2102.04788.
	
	\bibitem{Gravitational_fu_2021}
	Q. Fu and X. Zhang,
	Gravitational Lensing by a Black Hole in Effective Loop Quantum Gravity,
	arXiv:2111.07223 [gr-qc].
	
	\bibitem{consistent_bouhmadi-lopez_2020}
	M. Bouhmadi-Lopez, S. Brahma, C. Chen, P. Chen, and D. Yeom,
	A Consistent Model of Non-Singular Schwarzschild Black Hole in Loop Quantum Gravity and Its Quasinormal Modes,
	J. Cosmol. Astropart.Phys. 2020, 066 (2020).
	\bibitem{Liu22}S. J. Yang, Y. P. Zhang, S. W. Wei and Y. X. Liu,
Destroying the event horizon of a nonsingular rotating quantum-corrected black hole, JHEP 04 (2022) 066, arXiv:2201.03381 [gr-qc].
	
	\bibitem{Testing_brahma_2021}
	S. Brahma, C. Chen, and D. Yeom,
	Testing Loop Quantum Gravity from Observational Consequences of Nonsingular Rotating Black Holes,
	Phys. Rev. Lett. 126, 181301 (2021)
	
	
	\bibitem{Observational_zhu_2020}
	T. Zhu and A. Wang,
	Observational tests of the self-dual spacetime in loop quantum gravity,
	Phys. Rev. D 102, 124042 (2020).
	
	
	\bibitem{PROGRESS_fomalont_2009}
	E. Fomalont, S. Kopeikin, G. Lanyi, and J. Benson,
	PROGRESS IN MEASUREMENTS OF THE GRAVITATIONAL BENDING OF RADIO WAVES USING THE VLBA,
	Astron. J. 699, 1395 (2009).
	
	\bibitem{test_bertotti_2003}
	B. Bertotti, L. Iess, and P. Tortora,
	A Test of General Relativity Using Radio Links with the Cassini Spacecraft,
	Nature 425, 374 (2003).
	
	\bibitem{Precession_park_2017}
	R. S. Park, W. M. Folkner, A. S. Konopliv, J. G. Williams, D. E. Smith, and M. T. Zuber,
	Precession of Mercury's Perihelion from Ranging to the MESSENGER Spacecraft,
	Astron. J. 153, 121 (2017).
	
	\bibitem{Accurate_lucchesi_2010}
	D. M. Lucchesi and R. Peron,
	Accurate Measurement in the Field of the Earth of the General-Relativistic Precession of the LAGEOS II Pericenter and New Constraints on Non-Newtonian Gravity,
	Phys. Rev. Lett. 105, 231103 (2010).
	
	
	\bibitem{Progress_williams_2004}
	J. G. Williams, S. G. Turyshev, and D. H. Boggs,
	Progress in Lunar Laser Ranging Tests of Relativistic Gravity,
	Phys. Rev. Lett. 93, 261101 (2004).
	
	\bibitem{Metric_johannsen_2011}
	T. Johannsen and D. Psaltis,
	Metric for Rapidly Spinning Black Holes Suitable for Strong-Field Tests of the No-Hair Theorem,
	Phys. Rev. D 83, 124015 (2011).
	
	\bibitem{New_rezzolla_2014}
	L. Rezzolla and A. Zhidenko,
	New Parametrization for Spherically Symmetric Black Holes in Metric Theories of Gravity,
	Phys. Rev. D 90, 084009 (2014).
	
	\bibitem{BepiColombo_benkhoff_2010}
	J. Benkhoff, J. van Casteren, H. Hayakawa, M. Fujimoto, and et al.,
	BepiColombo \textemdash  Comprehensive Exploration of Mercury: Mission Overview and Science Goals,
	Planet Space Sci. 58, 2(2010).
	
	\bibitem{New_will_2018}
	C. M. Will,
	New General Relativistic Contribution to Mercury's Perihelion Advance,
	Phys. Rev. Lett. 120, 191101 (2018).

	\bibitem{Gravity_will_2016}
	C. M. Will,
	Gravity: Newtonian, Post-Newtonian, and General Relativistic, in Gravity: Where Do We Stand?, edited by R. Peron, M. Colpi, V. Gorini, and U. Moschella (Springer International Publishing, Cham, 2016).
	
	
\end{thebibliography}

\end{document}